\documentstyle[aps,12pt]{revtex}

\begin{document}
\draft
\author{O. B. Zaslavskii}
\address{Department of Physics, Kharkov State University, Svobody Sq.4, Kharkov\\
310077, Ukraine\\
E-mail: aptm@kharkov.ua}
\title{Semi-infinite throats at finite temperature and static solutions in exactly
solvable models of 2d dilaton gravity}
\maketitle

\begin{abstract}
Found is a general form of static solutions in exactly solvable models of 2d
dilaton gravity at finite temperature. We reveal a possibility for the
existence of everywhere regular solutions including black holes,
semi-infinite throats and star-like configurations. In particular, we
consider the Bose-Parker-Peleg (BPP) model which possesses a semi-infinite
throat and analyze it at finite temperature. We also suggest generalization
of the BPP model in which the appearance of semi-infinite throat has a
generic character and does not need special fine tuning between parameters
of the solution.
\end{abstract}

\pacs{PACS number(s): 04.60.Kz, 04.70.Dy}

\section{introduction}

Hawking's discovery of black hole evaporation posed the question about the
final fate of a black hole. This issue is far from being fully understood
but exploiting two-dimensional (2d) models gained insight into this
intriguing puzzle. In particular, Bose, Parker and Peleg (BPP) suggested the
exactly solvable modification of the Callan-Giddings-Harvey-Strominger
(CGHS) model \cite{callan} in which a black hole evaporates leaving, as an
end state, a regular static semi-infinite throat \cite{bose}. The existence
of such a type of solution motivates an interest to finding regular static
solutions in 2d dilaton gravity in a more general setting.

Apart from the possible role of such solutions as ''remnants'' after the
evaporation of black hole, this task is also motivated by the necessity to
elucidate the structure of exactly solvable solutions (of any types) in
dilaton gravity. In the paper \cite{kaz} the approach was suggested which
was based on a treatment of non linear $\sigma $ model. A more simple
approach appealing directly to the properties of the action coefficient as
functions of a dilaton field was proposed in \cite{zasl99} where it was also
shown that the general structure of found exactly solvable models
encompasses all known particular ones. The paper \cite{zasl99} deals only
with black holes with a regular horizon. Correspondingly, the temperature is
put equal to the Hawking one. In the present paper we consider another types
of a metric and find the general form of static solutions in exactly
solvable models at an arbitrary temperature.

We carry out the analysis of the obtained general solutions for the BPP
model and generalizations of it which, as well as BPP\ ones, do not contain
singularities at finite values of a dilaton field. It turns out that in
these models there exists a rather rich family of solutions regular
everywhere. In spite of the term ''throat''\ in 1+1 dimensional world is
somewhat conditional, we will use it, following \cite{bose}, to denote any
regular geodesically complete geometry extending to infinity (which for
definiteness is supposed to be a left one) where its curvature is nonzero.
If a metric is flat there, we will speak about a soliton-like configuration.
It is supposed that at the right infinity a spacetime is flat in any case.
We will see that, depending on temperature, the found family of exact
solutions contains regular spacetimes of all three possible types - black
holes, throats or soliton-like configurations.

\section{general form of static solutions}

Consider the action 
\begin{equation}
I=I_{0}+I_{PL}\text{,}  \label{action}
\end{equation}
where 
\begin{equation}
I_{0}=\frac{1}{2\pi }\int_{M}d^{2}x\sqrt{-g}[F(\phi )R+V(\phi )(\nabla \phi
)^{2}+U(\phi )]  \label{clac}
\end{equation}
and the Polyakov-Liouville action \cite{pl} incorporating effects of Hawking
radiation and its back reaction on the black hole metric can be written as 
\begin{equation}
I_{PL}=-\frac{\kappa }{2\pi }\int_{M}d^{2}x\sqrt{-g}[\frac{(\nabla \psi )^{2}%
}{2}+\psi R]  \label{PL}
\end{equation}
The function $\psi $ obeys the equation 
\begin{equation}
\Box \psi =R  \label{psai}
\end{equation}
where $\Box =\nabla _{\mu }\nabla ^{\mu }$, $\kappa =N/24$ is the quantum
coupling parameter, $N$ is number of scalar massless fields, $R$ is a
Riemann curvature. We omit the boundary terms in the action as we are
interested only in field equations and their solutions.

Varying the action with respect to a metric gives us $(T_{\mu \nu }=2\frac{%
\delta I}{\delta g^{\mu \nu }})$: 
\begin{equation}
T_{\mu \nu }\equiv T_{\mu \nu }^{(0)}+T_{\mu \nu }^{(PL)}=0  \label{6}
\end{equation}
where 
\begin{equation}
T_{\mu \nu }^{(0)}=\frac{1}{2\pi }\{2(g_{\mu \nu }\Box F-\nabla _{\mu
}\nabla _{\nu }F)-Ug_{\mu \nu }+2V\nabla _{\mu }\phi \nabla _{\nu }\phi
-g_{\mu \nu }V(\nabla \phi )^{2}\}\text{,}  \label{7}
\end{equation}
\begin{equation}
T_{\mu \nu }^{(PL)}=-\frac{\kappa }{2\pi }\{\partial _{\mu }\psi \partial
_{\nu }\psi -2\nabla _{\mu }\nabla _{\nu }\psi +g_{\mu \nu }[2R-\frac{1}{2}%
(\nabla \psi )^{2}]\}  \label{8}
\end{equation}

Variation of the action with respect to $\phi $ gives rise to the equation 
\begin{equation}
R\frac{dF}{d\phi }+\frac{dU}{d\phi }=2V\Box \phi +\frac{dV}{d\phi }(\nabla
\phi )^{2}  \label{9}
\end{equation}

Hereafter we assume that a dilaton field is not constant identically (such
special kinds of solutions are considered in \cite{solod}, \cite{zasl98}).
Then, as is shown in \cite{kaz}, \cite{zasl99}, we can gain a rich set of
exactly solvable models if the action coefficients in (\ref{clac}) $F(\phi )$%
, $V(\phi )$ and $U(\phi )\equiv 4\lambda ^{2}e^{\int_{0}^{\phi }\omega
(\phi ^{\prime })d\phi ^{\prime }}$are restricted by one constraint
equation. This equation can be written as 
\begin{equation}
V=\omega (u-\frac{\kappa \omega }{2})  \label{v}
\end{equation}
where $u\equiv \frac{dF}{d\phi }$. (In fact, eq. (\ref{v}) admits further
generalization due to a possible term $A(u-\kappa \omega )^{2}$ but, as it
leads, generally speaking, to metrics which are not asymptotically flat,
hereafter we put $A=0$). Then it turns out that the metric is a pure static.
In the present paper we will use the conformal gauge 
\begin{equation}
ds^{2}=g(-dt^{2}+d\sigma ^{2})  \label{conf}
\end{equation}
where $g=g(\sigma )$. In this gauge the curvature $R=-g^{-1}(g^{\prime
}/g)^{\prime }$. Throughout the paper the prime denotes differentiation with
respect to a spatial coordinate. Substituting it into (\ref{psai}) we get
after simple manipulations 
\begin{equation}
g=e^{-\psi -a\sigma }  \label{g}
\end{equation}
where $a$ is a constant. It is implied that time is normalized in such a way
that in (\ref{g}) the function $g\rightarrow 1$ at $\sigma \rightarrow
\infty $.

After simple rearrangement the (00) and (11) field equations (\ref{6}) with
the metric (\ref{g}) are reduced to one equation 
\begin{equation}
\xi _{1}\phi ^{\prime \prime }+\xi _{2}\phi ^{\prime 2}-\xi _{1}\frac{%
g^{\prime }\phi ^{\prime }}{g}=0  \label{field}
\end{equation}
where $\xi _{1}=\frac{d\tilde{F}}{d\phi }$, $\xi _{2}=\frac{d^{2}\tilde{F}}{%
d\phi ^{2}}-\tilde{V}$, $\tilde{F}=F-\kappa \psi $, $\tilde{V}=V-\frac{%
\kappa }{2}(\frac{d\psi }{d\phi })^{2}$.

In \cite{zasl99} we found that for exactly solvable models (\ref{v}) the
function $\psi $ is equal to $\psi \equiv \psi _{0}(\phi )$ where 
\begin{equation}
\psi _{0}=\int \omega d\phi  \label{ps0}
\end{equation}

For a given metric this function is defined up to an arbitrary function $%
\chi $ such that $\Box \chi =0$ as it follows from (\ref{psai}). For a
static metric (\ref{conf}) this gives us (up to the constant) $\chi =\gamma
\sigma $ where $\gamma $ is a constant. We will see below that for $\gamma
=0 $ we return to black hole solutions obtained in \cite{zasl99} whereas for 
$\gamma \neq 0$ we find new types of static solutions which include, in
particular, finite temperature generalizations of ''semi-infinite throats'' 
\cite{bose}. In what follows we will deal with solutions of the form

\begin{equation}
\psi =\psi _{0}+\gamma \sigma  \label{ps}
\end{equation}
with an arbitrary $\gamma \neq 0$. At the right infinity where spacetime is
supposed to be flat the function $\psi \sim \sigma \rightarrow \infty $ for
a generic $g$ \cite{sol95} irrespectively of whether or not $\gamma =0$.
Meanwhile, at the left coordinate infinity $\sigma \rightarrow -\infty $ (a
reader should bear in mind that if a spacetime is not geodesically complete,
the coordinate $\sigma $ does not cover the whole manifold) the admittance
of $\gamma \neq 0$ can qualitatively change the character of solution. For
instance, instead of a black hole for which at a horizon $g\rightarrow 0$
and $\psi $ is finite \cite{solod} one can get at a left infinity a
''semi-infinite throat'' \cite{bose} for which both $g$ and $\psi $ diverge
there to ensure, by definition of the ''semi-infinite throat'', the
divergencies of a proper distance at a finite value of a coordinate $\sigma $
(see below for details).

The structure of the action coefficients in our case is the same as in \cite
{zasl99}. Namely, the relationship between them is represented by eq. (\ref
{v}). The new feature introduced in our treatment as compared to \cite
{zasl99} is connected with another character of boundary conditions for the
function $\psi .$ Whereas in \cite{solod}, \cite{zasl99} one deals with a
black hole in the Hartle-Hawking state for which $\psi =\psi _{0}$ is
regular on a horizon, for solutions discussed below $\psi $ may diverge at $%
\sigma \rightarrow -\infty $ due to the term $\gamma \sigma $ in (\ref{ps}).

It is convenient to split coefficients in eq. (\ref{field}) into two parts
singling out the term which is built up with the help of $\psi _{0}$: $\xi
_{1}=\xi _{1}^{(0)}-\kappa \gamma \frac{d\sigma }{d\phi }$, $\xi _{2}=\xi
_{2}^{(0)}-\kappa \gamma \frac{d^{2}\sigma }{d\phi ^{2}}+\kappa [\frac{d\psi
_{0}}{d\phi }\gamma \frac{d\sigma }{d\phi }+\frac{1}{2}(\gamma \frac{d\sigma 
}{d\phi })^{2}]$, 
\begin{equation}
\xi _{1}^{(0)}=\frac{d\tilde{F}^{(0)}}{d\phi },\xi _{2}^{(0)}=\frac{d^{2}%
\tilde{F}^{(0)}}{d\phi ^{2}}-\tilde{V}^{(0)},\tilde{F}^{(0)}=F-\kappa \psi
_{0},\tilde{V}^{(0)}=V-\frac{\kappa }{2}(\frac{d\psi _{0}}{d\phi })^{2}.
\label{0}
\end{equation}

Then eq. (\ref{field}) takes the form 
\begin{equation}
\xi _{1}^{(0)}\phi ^{\prime \prime }+\xi _{2}^{(0)}\phi ^{\prime 2}+\xi
_{1}^{(0)}\phi ^{\prime }(\psi ^{(0)^{\prime }}+\delta )=\kappa \gamma
(\delta -\frac{\gamma }{2})  \label{f2}
\end{equation}
where $\delta =\gamma +a$. Let us multiply this equation by the factor $\eta 
$ such that $\xi _{2}^{(0)}\eta =\frac{d(\xi _{1}^{(0)}\eta )}{d\phi }$.
Then eq. (\ref{f2}) can be cast into the form 
\begin{equation}
z^{\prime }+z(\psi ^{(0)^{\prime }}+\delta )=\kappa \gamma (\delta -\frac{%
\gamma }{2})\eta  \label{y}
\end{equation}
where $z=\eta \xi _{1}^{(0)}\phi ^{\prime }=\eta \frac{d\tilde{F}}{d\sigma }%
^{(0)}$. It follows from (\ref{v}), (\ref{ps0}) and (\ref{0}) that $\eta
=e^{-\psi _{0}}$. Then after integration we get from (\ref{y}) 
\begin{equation}
\tilde{F}^{(0)}=C+De^{-\sigma \delta }+\kappa \gamma (1-\frac{\gamma }{%
2\delta })\sigma  \label{F}
\end{equation}
It follows from the trace of eq. (\ref{7}), (\ref{8}) that 
\begin{equation}
U=\Box \tilde{F}=\Box \tilde{F}^{(0)}  \label{pot}
\end{equation}
As for any function $f(\sigma )$ we have in the metric (\ref{g}) $\Box
f=g^{-1}f^{\prime \prime }$, we get from (\ref{pot}), (\ref{g}) and (\ref
{ps0}) the relationship between constants $D\delta ^{2}=4\lambda ^{2}$. In
what follows we assume that the function $\tilde{F}$ as well as $F$ has the
asymptotic behavior $\tilde{F}\simeq e^{\omega _{0}\phi }$ ($\omega
_{0}=const$) at the right infinity $\sigma \rightarrow \infty $ where $\phi
\rightarrow -\infty $, $\psi _{0}\simeq \omega _{0}\phi $ and the metric is
flat, $g\simeq e^{-\omega _{0}\phi -\delta \sigma }\rightarrow 1$. This
asymptotic condition is satisfied, in particular, by CGHS \cite{callan}, RST 
\cite{rst} and BPP models with $\omega _{0}=-2$. Thus, we get $D=1$, $\delta
=-2\lambda $ (it is assumed for definiteness that $\omega _{0}<0$) and the
relationship between a dilaton field and coordinate reads $\tilde{F}%
^{(0)}=f(y)\equiv e^{2y}-By+C$ where $y\equiv \lambda \sigma $, $B=-\kappa 
\frac{\gamma }{\lambda }(1+\frac{\gamma }{4\lambda })$. The value of $\gamma 
$ can be found from asymptotical conditions at right infinity where
spacetime is supposed to be flat. Let us impose such a condition which
describes quantum fields at finite temperature $T$. It means that at right
infinity the quantum stress-energy tensor should have the form 
\begin{mathletters}
\begin{equation}
T_{\mu }^{\nu (PL)}=\frac{\pi ^{2}N}{6}T^{2}(1,-1)  \label{T}
\end{equation}
Comparing it with the asymptotic form of eq. (\ref{8}), we find that $T_{\mu
}^{\nu (PL)}=\frac{\kappa }{4\pi }\psi _{\infty }^{\prime 2}(1,-1)$ where $%
\psi _{\infty }^{\prime }=\lim_{\sigma \rightarrow \infty }\frac{d\psi }{%
d\sigma }$. Hence, $\psi _{\infty }^{\prime 2}=\frac{2\pi ^{2}T^{2}N}{%
3\kappa }=16\pi ^{2}T^{2}$. Remembering that for exactly solvable models
under consideration the Hawking temperature is equal to $T_{0}=\lambda /2\pi 
$ \cite{zasl99} we obtain $\psi _{\infty }^{\prime 2}=4\lambda ^{2}\frac{%
T^{2}}{T_{0}^{2}}$. On the other hand, we have from (\ref{g}) and the
condition $g\rightarrow 1$ at infinity that $\psi _{\infty }^{\prime }=-a$.
Then $\gamma =\delta -a=-2\lambda +\psi _{\infty }^{\prime }$ $=2\lambda
(T/T_{0}-1)$ whence $B=\kappa (1-T^{2}/T_{0}^{2})$ (the sign of $\psi
_{\infty }^{\prime }$ is chosen to ensure $\gamma =0$ for $T=T_{0}$ when $%
\psi =\psi _{0}$ and we return to the situation described in \cite{zasl99}).
Collecting all basic formulas, we obtain 
\end{mathletters}
\begin{eqnarray}
ds^{2} &=&g(-dt^{2}+d\sigma ^{2})\text{, }g=\exp (-\psi _{0}+2y)\text{, }%
y=\lambda \sigma \text{, }\psi _{0}=\int \omega d\phi \text{.}  \label{bas}
\\
\tilde{F}^{(0)}(\phi ) &=&f(y)\equiv e^{2y}-By+C\text{, }B=\kappa
(1-T^{2}/T_{0}^{2})\text{, }T_{0}=\lambda /2\pi \text{.}  \nonumber
\end{eqnarray}
If $T>T_{0}$, the coefficient $B>0$ and the function takes its minimum at
the point $y_{0}=\frac{1}{2}\ln B$ where $f(y_{0})=C-C^{*}$, 
\begin{equation}
C^{*}=-\frac{B}{2}(1+\ln \frac{2}{B})  \label{c}
\end{equation}
Let $T=T_{0}$, then the coefficients $\gamma =B=0$ whence we obtain from (%
\ref{bas}) $g=\exp (-\psi _{0})(\tilde{F}^{(0)}-C)$. If we identify $C=%
\tilde{F}(\phi _{h})$ we obtain black hole solutions with the horizon at $%
\phi =\phi _{h}$ analyzed in \cite{zasl99}. Then the equality of
temperatures simply means that the state of the Euclidean metric we deal
with is the Hartle-Hawking one. To make the comparison more clear, let us
pass from the conformal gauge (\ref{g}) to the Schwarzschild one $%
ds^{2}=-gdt^{2}+g^{-1}dx^{2}$used in \cite{zasl99}. This can be achieved by
the coordinate transformation $x=\int d\sigma g$. It follows from (\ref{bas}%
) that $x=(2\lambda )^{-1}\int d\phi e^{-\psi }\tilde{F}^{\prime }$ that
agrees with eq. (27) of \cite{zasl99}.

It is worth stressing, however, that the class of solutions (\ref{bas}) is
much more general than that found in \cite{zasl99}. It includes not only
black holes due to the parameter $B$ which describes a system at arbitrary
temperature when black holes with a regular horizon cannot exist. In what
follows we will use the notation $T_{0}=\lambda /2\pi $ even in cases when $%
T_{0}$ does not have the meaning of the Hawking temperature. Equations (\ref
{bas}) represent in the closed form the general expression describing static
solutions in exactly solvable models of 2d gravity \cite{kaz}, \cite{zasl99}%
. The function in the right hand side of eq. (\ref{bas}) has the universal
form while a particular model is characterized by the choice of $\tilde{F}%
^{(0)}(\phi )$. It is the analysis of this class of solutions that we now
turn to. As we are mainly interested in finding regular solutions, we
discuss the BPP model and its generalizations.

\section{BPP model at finite temperature}

This model is specified by the choice 
\begin{equation}
F=e^{-2\phi }-2\kappa \phi \text{, }\tilde{F}=e^{-2\phi }\text{, }\omega =-2%
\text{, }\psi _{0}=-2\phi \text{, }V=4e^{-2\phi }+2\kappa  \label{bpp}
\end{equation}
If $T=T_{0}$ we have $B=0$ and $e^{2y}=e^{-2\phi }-e^{-2\phi _{h}}$, $%
g=1-e^{2(\phi -\phi _{h})}$. As seen from the above formulas, the expression
for the metric (either in terms of $\phi $ or in terms of a coordinate) has
the same form as in the classical limit ($\kappa =0$), so in this sense the
metric does not acquire quantum corrections \cite{zasl99}. In the paper \cite
{bose} the authors considered the case $T=0$. Then asymptotically the energy
density of quantum field approaches zero (radiationless solutions). In
general, spacetime contains a time-like singularity or a singular horizon
depending on the value of $C$ that determines whether or not the function $%
f(y)$ has a zero. The special case arises when simultaneously $%
f(y)=0=f^{\prime }(y)$ at $y=y_{0}$, $C=C^{*}=-\frac{\kappa }{2}(1+\ln \frac{%
2}{\kappa })$ \cite{bose} (a reader should have in mind that we define the
quantum parameter $\kappa =N/24$ whereas in \cite{bose} $\kappa =N/12$).
Then the spacetime does not contain singularities at all extending from the
left infinity where $R=-4\lambda ^{2}$ (semi-infinite throat) to the right
one where $R=0$.

In fact, the analysis is tractable to the case of an arbitrary $T$ due to
the universality and simplicity of the function $f(y)$ with an arbitrary $B$%
. Let, first, $T<T_{0}$, so $B>0$. For $C<C^{*}$ there is a time-like naked
singularity $(y=y_{0}$) at a finite proper distance. If $C>C^{*}$, there
exists a singular horizon at $\phi =\infty $. The throat exists provided $%
C=C^{*}$ where $C^{*}$, according to (\ref{c}), depends on temperature via $%
B(T)$. The calculation of curvature shows that at the throat $R=-4\lambda
^{2}$ independently of temperature. In the limit $T=0$ $B=\kappa $ and we
return to radiationless solutions described in \cite{bose}.

If $T>T_{0}$, $B<0$ the function $f(y)$ is monotonic and does not turn into
a zero. Then for any $C$ there is a time-like singularity at $\phi =\infty $.

In the point $T=T_{0}$ the coefficients $B=0=C^{*}$. If $C>0$ we have a
black hole in the Hartle-Hawking state with a regular horizon. If $C<0$
there exists a time-like singularity. For $C=0$ the spacetime is flat
(linear dilaton vacuum $\phi =-y$). The situation can be summarized as
follows.

BPP model

\begin{tabular}{|l|l|l|l|}
\hline
& $T<T_{0}$ & $T=T_{0}$, $B=0=C^{*}$ & $T>T_{0}$ \\ \hline
$C<C^{*}$ & time-like singularity & time-like singularity & time-like \\ 
\hline
$C=C^{*}$ & throat & linear dilaton vacuum, flat spacetime & singularity \\ 
\hline
$C>C^{*}$ & singular horizon & space-like singularity behind a horizon & for
any $C$ \\ \hline
\end{tabular}

The main feature of the model in question is the fact that the semi-infinite
throat may exist at any finite temperature $T<T_{0}$. Meanwhile, this needs
the special choice $C=C^{*}(T)$. It turns out, however, that there are
another exactly solvable models for which the existence of a throat is
generic and does need fine tuning in constants.

\section{generalizations of bpp model}

Consider the model 
\begin{eqnarray}
F^{(0)} &=&\alpha e^{-2b\phi }+e^{-2\phi }-2\kappa \phi \text{, }\alpha >0%
\text{, }0<b\leq 1/2\text{, }\omega =-2\text{, }V=4(e^{-2\phi }+b\alpha
e^{-\phi })+2\kappa \text{,}  \label{gen} \\
\tilde{F}^{(0)} &=&\alpha e^{-2b\phi }+e^{-2\phi }  \nonumber
\end{eqnarray}
which is, according to (\ref{v}), exactly solvable and the Hawking
temperature for black hole solutions is equal to $T_{0}$ \cite{zasl99}. Let
first $B>0$ ($T<T_{0}$). Then near the point $y_{0}$ where $f(y_{0})=0$ we
have for any $C<C^{*}$ $f\sim y-y_{0}$, $\tilde{F}^{(0)}\sim e^{-2b\phi
}\rightarrow 0$, $\phi \rightarrow \infty $, so $e^{2\phi }$ $\sim
(y-y_{0})^{-1/b}$ and, according to (\ref{bas}), $g=e^{2\phi +2y}\sim
(y-y_{0})^{-1/b}$. Then the curvature $R\sim (y-y_{0})^{1/b-2}$. If $b<1/2$,
the curvature $R=0$ at left infinity, the proper distance $l$ between $y_{0}$
and any other point $y>y_{0}$ diverges like $l\sim (y-y_{0})^{1-1/2b}$.
Thus, we obtain a soliton-like configuration.

If $C=C^{*}$, $f\sim (y-y_{0})^{2}$, $g$ $\sim (y-y_{0})^{-2/b}$, $R\sim
(y-y_{0})^{2/b-2}\rightarrow 0$, so the configuration is again soliton-like
for any $0<b<1$. For $C>C^{*}$ spacetime cannot be regular everywhere, it
contains the singular horizon where $y\rightarrow -\infty $, $g\sim
e^{2y}\left| y\right| ^{-1/b}$, $R\sim -e^{-2y}\left| y\right| ^{1/b-2}$, a
proper distance $l$ is finite.

The rest of cases can be treated in a similar manner. We only dwell on the
fact that for $T=T_{0}$ a spacetime represent a black hole whose curvature
is finite everywhere. In particular, at left infinity where $\phi
\rightarrow \infty $ the curvature $R\rightarrow 0$, so spacetime is
asymptotically flat at both infinities. The variety of cases is tabulated as
follows.

$\tilde{F}^{(0)}=e^{-2\phi }+\alpha e^{-2b\phi }$, $0<b<1/2$

\begin{tabular}{|l|l|l|l|}
\hline
& $T<T_{0}$ & $T=T_{0}$, $B=0=C^{*}$ & $T>T_{0}$ \\ \hline
$C<C^{*}$ & soliton-like & soliton-like & soliton-like \\ \hline
$C=C^{*}$ & soliton-like for $0<b<1$ & soliton-like & for \\ \hline
$C>C^{*}$ & singular horizon & black hole regular everywhere & any $C$ \\ 
\hline
\end{tabular}

Let now $b=1/2$. The results of consideration, details of which we omit, can
be summarized in the table below.

$\tilde{F}^{(0)}=e^{-2\phi }+\alpha e^{-\phi }$

\begin{tabular}{|l|l|l|l|}
\hline
& $T<T_{0}$ & $T=T_{0}$, $B=0=C^{*}$ & $T>T_{0}$ \\ \hline
$C<C^{*}$ & throat & throat & throat \\ \hline
$C=C^{*}$ & soliton-like & soliton-like & for \\ \hline
$C>C^{*}$ & singular horizon & black hole regular everywhere & any $C$ \\ 
\hline
\end{tabular}

The most interesting feature of this model is the existence of a throat at
any temperature (for $C<C^{*}$, if $T\leq T_{0}$ and for an arbitrary $C$ if 
$T>T_{0}$).

\section{Discussion and conclusion}

We have found a general form of static solutions in 2d exactly solvable
dilaton gravity theories accounting for quantum effects at finite
temperature. The criteria of solvability are the same as derived in \cite
{kaz}, \cite{zasl99} for eternal black holes and represent one constraint on
the action coefficient (\ref{v}). Thus, every model with black hole
solutions considered in \cite{zasl99} generates a family of solutions of
another kind which generalize \cite{zasl99} and reduce to it in the
particular case $T=T_{0}$ when the solution describes a black hole with a
Hawking temperature $T_{0}$ in thermal equilibrium with its Hawking
radiation. It was indicated in \cite{zasl99} that exactly solvable models in
question exhibits a series of universal properties. In particular, it is
remarkable that a Hawking temperature is constant for all such models, $%
T_{0}=\lambda /2\pi $. The present consideration extended this universality
to a much more wide class solution and enabled us to classified them in an
unified manner due to the universal structure of (\ref{bas}) independently
of the choice of the model. In fact, the particular properties of the model
are encoded in the only function $\tilde{F}^{(0)}(\phi )$. If $\tilde{F}%
^{(0)\prime }(\phi )$ does not have zeros at finite $\phi $, as it takes
place for the BPP model \cite{bose} and its generalization considered in the
present paper, the only relevant information is contained in the asymptotic
behavior of $\tilde{F}^{(0)}$ at $\phi \rightarrow \infty $. We analyzed two
concrete examples and found that there exist three types of solutions
regular everywhere. First, it is a black hole, if $T=T_{0}$ (the possibility
of solutions of such a kind was pointed out in \cite{zasl99}). Second, this
a semi-infinite throat generalizing the observation made in \cite{bose}. The
most interesting feature of throats found in the model (\ref{gen}) with $%
b=1/2$ is that such types of solutions, unlike the BPP case, do not need a
special choice of parameters of the solution:\ semi-infinite throats may
exist in the whole range of parameters and temperatures. Third case is a
soliton-like configuration extending in both directions to flat infinities
without horizons. Such a type of solution is absent in the BPP model but may
occur in the generalizations of it.

The conditions which make the existence of semi-infinite throats possible
are essentially quantum. Indeed, as seen from (\ref{bas}), such a kind of
configuration arises only if the function $f(y)$ is not monotonic due to the
coefficient $B$ and takes its minimum value at some finite $y_{0}=\frac{1}{2}%
\ln B$. However, in the classical limit $\kappa \rightarrow 0$ the
coefficient $B\rightarrow 0$ and we get $g=e^{-\psi _{0}}(F^{(0)}-C)$.
Choosing $C=F(\phi _{h})$ we can get a classical counterpart of a quantum
black hole but not a semi-infinite throat. For example, with $\omega =-2$, $%
F^{(0)}=e^{-2\phi }$ we have the familiar expression for a static dilaton
black hole $g=1-Ce^{2\phi }$ where $C/2$ plays the role of a classical ADM
mass.

In the present paper we determined possible types of static solutions but
did not consider dynamical scenarios in the process of which they could
arise. Here we only restrict ourselves by a remark that finite temperature
solutions allow us to consider more complicated processes of interaction
between a black hole and its quantum surrounding than a evaporation into
vacuum. In particular, this enables us to include into consideration
directly the role of thermal bath \cite{bath}. Another interested problem
for further researches is looking for possibilities to gain new exactly
solvable models not only due to relaxing the regularity condition for the
function $\psi $ at the left infinity (as was made in the present paper) but
also to the generalization of the action structure itself. All this needs a
special treatment.

I\ am grateful to Sergey Solodukhin for careful reading the manuscript and
valuable comments. This work is supported by International Science Education
Program (ISEP), grant No. QSU082068.





%
%

%
%

\end{document}